\documentclass[aps,prl,twocolumn,superscriptaddress,
preprintnumbers,
amsmath,amssymb,showkeys,floatfix,nofootinbib,reprint]{revtex4-1}

\usepackage{amsmath}
\usepackage{amsfonts}
\usepackage{amssymb}
\usepackage{graphicx}
\usepackage{color}
\usepackage{bm}
\usepackage{hyperref}

\begin{document}

%
%
% -------------------------------- our notations --------------------------- %
%

\def\ada#1{\textcolor{blue}{#1}}
\def\jonas#1{\textcolor{red}{#1}}

\def\ket#1{ $ \left\vert  #1   \right\rangle $}
\def\ketm#1{  \left\vert  #1   \right\rangle   }
\def\bra#1{ $ \left\langle  #1   \right\vert $ }
\def\bram#1{  \left\langle  #1   \right\vert   }
\def\spr#1#2{ $ \left\langle #1 \left\vert \right. #2 \right\rangle $ }
\def\sprm#1#2{  \left\langle #1 \left\vert \right. #2 \right\rangle   }
\def\me#1#2#3{ $ \left\langle #1 \left\vert  #2 \right\vert #3 \right\rangle $}
\def\mem#1#2#3{  \left\langle #1 \left\vert  #2 \right\vert #3 \right\rangle   }
\def\redme#1#2#3{ $ \left\langle #1 \left\Vert
                  #2 \right\Vert #3 \right\rangle $ }
\def\redmem#1#2#3{  \left\langle #1 \left\Vert
                  #2 \right\Vert #3 \right\rangle   }
\def\threej#1#2#3#4#5#6{ $ \left( \matrix{ #1 & #2 & #3  \cr
                                           #4 & #5 & #6  } \right) $ }
\def\threejm#1#2#3#4#5#6{  \left( \matrix{ #1 & #2 & #3  \cr
                                           #4 & #5 & #6  } \right)   }
\def\sixj#1#2#3#4#5#6{ $ \left\{ \matrix{ #1 & #2 & #3  \cr
                                          #4 & #5 & #6  } \right\} $ }
\def\sixjm#1#2#3#4#5#6{  \left\{ \matrix{ #1 & #2 & #3  \cr
                                          #4 & #5 & #6  } \right\} }

\def\ninejm#1#2#3#4#5#6#7#8#9{  \left\{ \matrix{ #1 & #2 & #3  \cr
                                                 #4 & #5 & #6  \cr
                         #7 & #8 & #9  } \right\}   }
%
%
% ---------------------------- end of our notations --------------------------- %
%
%

%
% -----------------------------------------    Title of paper ---------------------------------------
%

\title{Tailoring laser-generated plasmas for efficient nuclear excitation by electron capture}

%
% ------------------------------------------   List of authors --------------------------------------
%

\author{Yuanbin \surname{Wu}}
\email{yuanbin.wu@mpi-hd.mpg.de}
\affiliation{Max-Planck-Institut f\"ur Kernphysik, Saupfercheckweg 1, D-69117 Heidelberg, Germany}

\author{Jonas \surname{Gunst}}
\email{Jonas.Gunst@mpi-hd.mpg.de}
\affiliation{Max-Planck-Institut f\"ur Kernphysik, Saupfercheckweg 1, D-69117 Heidelberg, Germany}

\author{Christoph H. \surname{Keitel}}
\affiliation{Max-Planck-Institut f\"ur Kernphysik, Saupfercheckweg 1, D-69117 Heidelberg, Germany}

\author{Adriana \surname{P\'alffy}}
\email{Palffy@mpi-hd.mpg.de}
\affiliation{Max-Planck-Institut f\"ur Kernphysik, Saupfercheckweg 1, D-69117 Heidelberg, Germany}

%Date

\date{\today}

%
%
%
% ---------------------------------------------------- Abstract ---------------------------------------------
%
%
%
%
\begin{abstract}

The optimal parameters for nuclear excitation by electron capture in plasma environments generated  by the interaction of ultra-strong optical lasers with solid matter are investigated theoretically.
As a case study we consider a 4.85 keV nuclear transition starting from the long-lived $^{93\mathrm{m}}$Mo isomer that can lead to the release of the stored 2.4 MeV excitation energy. We find that due to the complex  plasma dynamics, the  nuclear excitation rate and the actual number of excited nuclei do not reach their maximum at the same laser parameters. The nuclear excitation achievable with a high-power optical laser is up to twelve and up to six orders of magnitude larger than the values predicted for  direct resonant and secondary plasma-mediated excitation at the x-ray free electron laser, respectively.  Our results show that the experimental observation of the nuclear excitation of $^{93\mathrm{m}}$Mo and the subsequent release of stored energy should be  possible at laser facilities available today.

\end{abstract}
%
%

% POSSIBILITIES FOR SUBMISSION, determines order and choice of PACS numbers!

% 1. Nuclear physics
% 2. Atomic, Molecular, and Optical Physics
% 3. Plasma and Beam Physics

% remark: plasma and atomic have significantly more articles in PRL than nuclear!


%23.20.Nx, % Internal conversion and extranuclear effects (including Auger electrons and internal bremsstrahlung)
%23.35.+g, % Isomer decay
%52.25.Os, % Emission, absorption, and scattering of electromagnetic radiation
%41.60.Cr % Free-electron lasers
%34.80.Lx, % Recombination, attachment, and positronium formation
%52.20.-j, % Elementary processes in plasmas
%52.70.La, % X-ray and gamma-ray measurements
%52.38.-r 	Laser-plasma interactions (for plasma production and heating by laser beams, see 52.50.Jm)
%52.50.Jm 	Plasma production and heating by laser beams (laser-foil, laser-cluster, etc.)

%\keywords{}

\maketitle

%--------------------------------------------------------------------------------------
%\section{Intro}

Novel coherent light sources open unprecedented possibilities for the field of laser-matter interactions \cite{DiPiazzaRMP2012}. The X-ray Free Electron Laser (XFEL) \cite{LCLS-web,Sacla} for instance can drive low-energy electromagnetic transitions in nuclei. Ultra-strong optical laser systems with up to few petawatt power \cite{ELI-BL-web,PETAL-web,CasnerHEDP2015,LULI-web,Vulcan-web} are very efficient in generating plasma environments  \cite{bookDieter}, which  host  complex interactions between photons, electrons, ions and the atomic nucleus. Nuclear excitation in laser-generated hot plasmas involving  optical lasers \cite{HarstonPRC1999, Gosselin2004, Gosselin2007, Morel2004-local, Morel2004-local-Hg, Morel2010-nonlocal, Comet2015-NEET, KenL2000.PRL, Cowan2000.PRL, Gibbon2005.Book, Spohr2008.NJP, Mourou2011.S,Andreev2000, Andreev2001, Granja2007-list-nuclei, Renner2008, Gobet2011}, or cold high-density plasmas \cite{Vinko2012.N} at the  XFEL  \cite{GunstPRL2014, GunstPOP2015} have been under investigation. Special attention has been attracted by nuclear transitions starting from long-lived excited states.
Such states are also known as nuclear isomers and are particularly interesting due to their potential to store large amounts of energy over long periods of time \cite{WalkerN1999, AprahamianNP2005, BelicPRL1999, CollinsPRL1999, BelicPRC2002, CarrollLPL2004, PalffyPRL2007,WalkerReview2016}. A typical example is  $^{93\mathrm{m}}$Mo at 2.4 MeV, for which an additional excitation of only 4.85~keV could lead to the depletion of the isomer and release on demand of the stored energy.

For both optical- and x-ray laser-generated plasmas, the process of nuclear excitation by electron capture (NEEC) \cite{GoldanskiiPLB1976, PalffyCP2010} into the atomic shell has proven to have a significant contribution. As secondary process in the cold plasma environment generated by the interaction of the XFEL with solid-state targets,  NEEC can exceed the direct nuclear photoexcitation  by six orders of magnitude    \cite{GunstPRL2014, GunstPOP2015} for the 4.85 keV excitation starting from the  $^{93\mathrm{m}}$Mo isomeric state. In this Letter, we show that by tailoring optical-laser-generated plasmas to harness maximum nuclear excitation via NEEC,  a further six orders of magnitude increase in the nuclear excitation and subsequent isomer depletion compared to the case of cold XFEL-generated plasmas  can be reached. As an interesting point, we find that due to the complexity of the processes involved, the plasma and correspondingly laser parameters for reaching the maximal NEEC rate are  not identical to the ones that provide the maximal number of nuclei actually excited. Our calculations demonstrate that the maximal number of depleted isomers for realistic laser setup parameters may reach for the first time measurable values. Although still far from the final goal, this is a further milestone on the way to the realization of controlled energy storage and release via nuclear isomers.

We consider a strong optical laser that interacts with a solid-state target containing a  fraction of nuclei in the isomeric state.  NEEC and photoexcitation  may occur in the generated plasma.
In the resonant process of NEEC, a free electron recombines into a vacant bound atomic state with the simultaneous excitation of the nucleus. The isomers can then be excited to a trigger state which rapidly decays to the nuclear ground state and releases the stored energy. We consider in the following the case of $^{93\mathrm{m}}$Mo for which recent claims have been made \cite{workshopGWU} on the first observation of NEEC following the proposals in Refs.~\cite{KaramianPAN2012, PolasikPRC2017}.

Free electrons in the plasma cover  a broad energy range such that many NEEC resonance channels may contribute to the net NEEC rate $\lambda_{\text{neec}}$.  This can be expressed as the convolution over the electron energy $E$ of the almost Dirac-delta-like NEEC single-resonance cross section $\sigma_{\text{neec}}$ and the free-electron flux $\phi_{\text{e}}$, summed over all charge states $q$ and capture channels $\alpha_{\text{d}}$,
\begin{equation} \label{eq:lambda}
    \lambda_{\text{neec}}(T_{\text{e}}, n_{\text{e}}) = \sum_{q, \alpha_{\text{d}}} P_q(T_{\text{e}}, n_{\text{e}})\! \int\! dE \, \sigma_{\text{neec}}(E) \phi_{\text{e}}(E,T_{\text{e}}, n_{\text{e}}).
\end{equation}
Here, $P_q$ is the probability to find ions charge state $q$ in the plasma as a function of electron temperature $T_{\text{e}}$ and density $n_{\text{e}}$. 
The dependence  $\phi_{\text{e}}(T_{\text{e}})$  determines the quantitative contribution of the NEEC resonances. The theoretical formalism for the calculation of the NEEC cross section $\sigma_{\text{neec}}$ has been presented elsewhere ~\cite{PalffyPRA2006, PalffyPRA2007, GunstPRL2014, GunstPOP2015}. The total NEEC excitation number $N_{\text{exc}}$ is connected to the rate $\lambda_{\text{neec}}$ via
\begin{equation} \label{eq:N_exc}
    N_{\text{exc}} = \int_{V_{\text{p}}}\! d^3\bold{r} \int\! dt\, n_{\text{iso}}(\bold{r},t) \, \lambda_{\text{neec}}(T_{\text{e}}, n_{\text{e}};\bold{r},t),
\end{equation}
where $n_{\text{iso}}$ denotes the number density of isomers and $V_{\text{p}}$ is the plasma volume.
 Let us assume in a first approximation homogeneous plasma conditions over the plasma lifetime $\tau_{\text{p}}$. Then the total number of excited nuclei is $N_{\text{exc}} = N_{\text{iso}} \lambda_{\text{neec}}(T_{\text{e}}, n_{\text{e}}) \tau_{\text{p}}$, with $N_{\text{iso}}$ the number of isomers in the plasma. Assuming a spherical plasma the plasma lifetime is approximatively given by 
$\tau_{\text{p}} = R_{\text{p}} \sqrt{m_{\text{i}}/(T_{\text{e}} \bar{Z})}$ \cite{KrainovS2002, GunstPOP2015}
with the ion mass $m_{\text{i}}$, the average charge state $\bar{Z}$ and the plasma radius $R_{\text{p}}$. $N_{\text{iso}}$ can be estimated introducing the isomer fraction embedded in the original solid-state target $f_{\text{iso}}$, $N_{\text{iso}} = f_{\text{iso}} n_{\text{i}} V_{\text{p}}$, where $n_{\text{i}}$ stands for the ion number density in the plasma.
A ${}^{93\text{m}}$Mo isomer fraction of $f_{\text{iso}} \approx  10^{-5}$ embedded in solid-state Niobium foils can be generated by intense ($\ge 10^{14}$ protons/s) beams  \cite{GunstPRL2014} via the $^{93}_{41}$Nb(p,n)$^{93\mathrm{m}}_{\phantom{m} 42}$Mo reaction \cite{exfor}.

\begin{figure}
\centering
\includegraphics[width=1.0\linewidth]{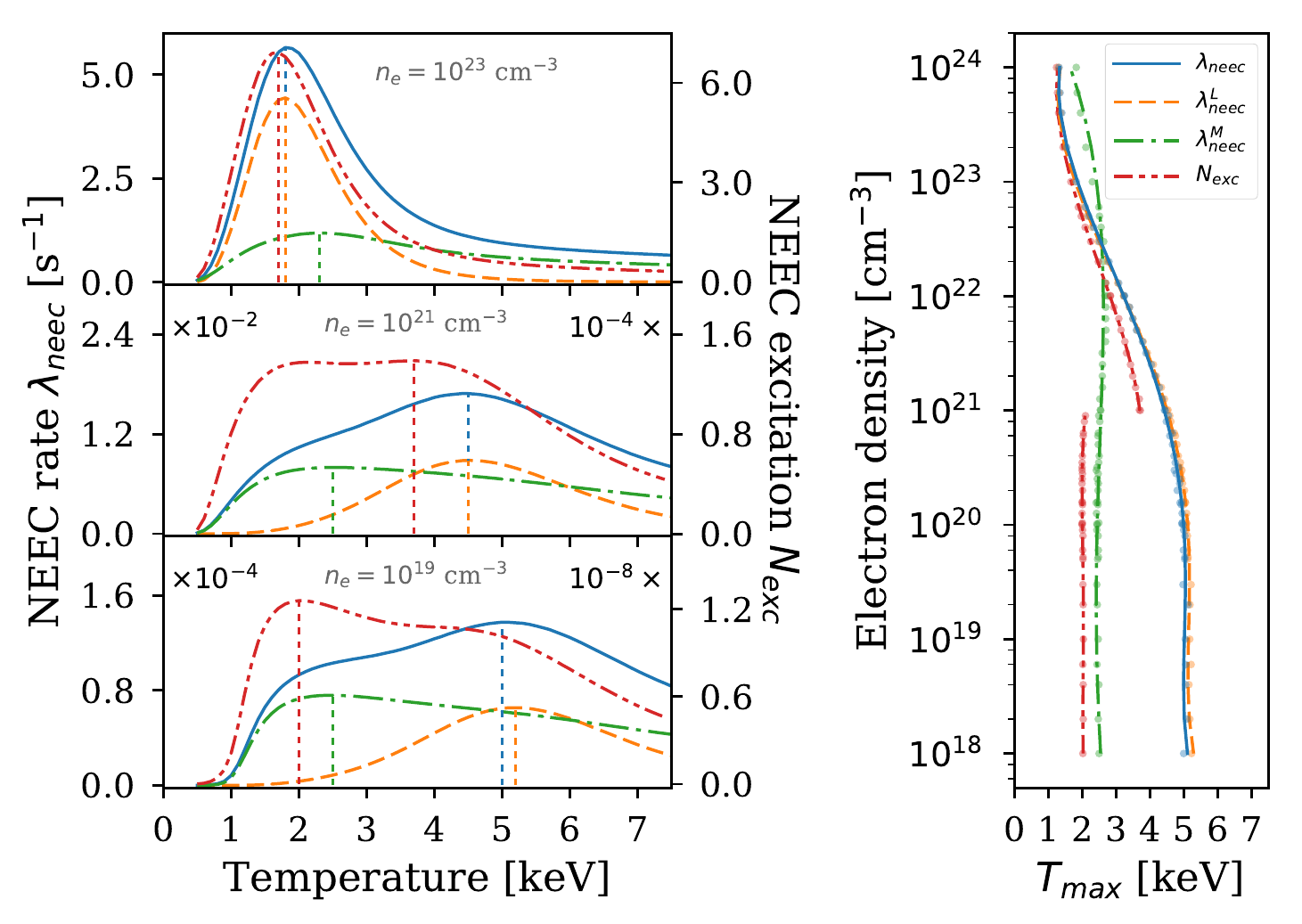}%
\caption{(color online). Left graph: NEEC rate $\lambda_{\text{neec}}$ (blue, solid curve) and the total number of excited isomers $N_{\text{exc}}$ (red, dash-dotted curve), as well as the individual contributions $\lambda_{\text{neec}}^{\text{L}}$ (orange, dashed curve) and $\lambda_{\text{neec}}^{\text{M}}$ (green, dash-dotted curve) from the $L$ and $M$ shell, respectively,  as a function of the electron temperature $T_{\text{e}}$ for selected electron densities $n_{\text{e}}$. A plasma radius of 40 $\mu$m has been assumed in the calculations of $N_{\text{exc}}$. Right graph:  Temperatures $T_{\text{max}}$ as functions of density, for maximizing $N_{\text{exc}}$, $\lambda_{\text{neec}}$, $\lambda_{\text{neec}}^{\text{L}}$ and $\lambda_{\text{neec}}^{\text{M}}$, respectively, at each particular $n_{\text{e}}$.
\label{fig:general}}
\end{figure}

Numerical results for $\lambda_{\text{neec}}$ and the corresponding total number of excited isomers $N_{\text{exc}}$  for an arbitrary plasma radius of 40 $\mu$m are presented in Fig.~\ref{fig:general}. For the calculation of $ \sigma_{\text{neec}}$ we use a theoretical prediction for the reduced nuclear transition probability \cite{Hasegawa2011.PLB}.  We model the plasma conditions by a relativistic distribution for the free electrons and a charge state distribution computed with the radiative-collisional code FLYCHK \cite{FLYCHK2005}. The relativistic electronic wave functions \cite{GRASP1989} and binding energies are in first approximation  calculated independently of $T_e$ and $n_e$, which are accounted for only indirectly via the charge state distribution. For a specific charge state, we further assume that the ion is in its ground state and recombination of the NEEC electron occurs in a free orbital. Among these assumptions, neglecting the plasma-induced ionization potential depression \cite{StPy1966}  is the most severe appoximation, as binding energies may vary by few eV to hundreds of eV depending on the plasma density. Using the Steward-Pyatt model \cite{StPy1966} which also has reasonable agreement with more recently developed methods \cite{LinPRE2017,HuPRL2017} for our purpose, we estimate that the consequences for $N_{\text{exc}}$   are even for the case of high-density plasmas with large ionization potential depression only on the level of ~10\%.

%%%%%%%%%%%%%%%%%%%%%%%%%%%%%%%%%%%%

%%%%%%%%%%%%%%%%%%%%%%%%%%%%%%%%%%%%%

NEEC into the $K$ shell is energetically forbidden for the 4.85 keV transition in Mo. The results for the dominant recombination channels  into the $L$ and $M$  atomic shells are presented individually in Fig.~\ref{fig:general}.
For the total NEEC rate $\lambda_{\text{neec}}$, further smaller contributions from the recombination into the $N$ and $O$ shells were also taken into account. Both  $\lambda_{\text{neec}}$ and $N_{\text{exc}}$ increase with increasing electron density $n_{\text{e}}$. In the range $n_{\text{e}} = 10^{19}$ cm$^{-3}$ to $10^{20}$ cm$^{-3}$, our calculations show that the charge state distribution $P_q$ is nearly unaffected for a fixed temperature $T_{\text{e}}$, while $\lambda_{\text{neec}}$ is enhanced by a factor of 10 maintaining the same functional dependence on $T_{\text{e}}$.
This  indicates that at low densities the boost in $\lambda_{\text{neec}}$ is (almost) a pure density effect coming from the increasing number of free electrons present in the plasma ($\phi_{\text{e}} \propto n_{\text{e}}$).
Increasing the electron density to even higher values, the behavior of $\lambda_{\text{neec}}$ and $N_{\text{exc}}$ becomes more involved as the charge distribution $P_q$ shows a complex dependence on the plasma conditions $n_{\text{e}}$ and $T_{\text{e}}$. Between $n_{\text{e}} = 10^{21}$ cm$^{-3}$ and $10^{23}$ cm$^{-3}$ we see that  with increasing $n_{\text{e}}$ the atomic shell contributions change significantly and $\lambda_{\text{neec}}$ is much enhanced.

The temperature $T_{\text{max}}$ at which $N_{\text{exc}}$ or the total or partial shell contributions $\lambda_{\text{neec}}$ reach a maximum for each density value $n_{\text{e}}$ is depicted in the right graph of Fig.~\ref{fig:general}. Naively, one would expect that $T_{\text{max}}$ is approximately the same for $N_{\text{exc}}$ and for $\lambda_{\text{neec}}$. This is however only true at high densities starting from $10^{21}$ cm$^{-3}$. According to our approximation for $\tau_{\text{p}}$, the chosen plasma lifetime is $T_{\text{e}}$-dependent. In particular at low electron densities, $\tau_{\text{p}}$ acts as a weighting function proportional to $(T_{\text{e}})^{-1/2}$ shifting the maximum of $N_{\text{exc}}$ to lower temperatures. The optimal plasma conditions for the total excitation number can thus drastically differ from the optimal conditions for $\lambda_{\text{neec}}$ in this model. We  note  that the arbitrary choice of $R_{\text{p}}$ only influences the absolute scale of the NEEC excitation number, not the position of $T_{\text{max}}$.

A comparison with nuclear photoexcitation assuming a black-body radiation spectrum at the given plasma temperature $T_{\text{e}}$ shows that at  $n_{\text{e}} = 10^{21}$ cm$^{-3}$ NEEC dominates for $T_{\text{e}}<1.6$ keV and  for higher densities $n_{\text{e}} = 10^{22}$ cm$^{-3}$ up to a temperature of 5 keV. The actual photoexcitation in the plasma should be even lower  in particular at low densities because photons may easier escape the finite plasma volume. 
For the high density $n_{\text{e}} \ge 10^{23}$ cm$^{-3}$ parameter regime, NEEC is the dominant nuclear excitation mechanism.

In the following we proceed to determine how the optimal NEEC parameter region in the temperature-density landscape may be accessed by a short laser pulse.  We discern in our treatment two cases, namely the low- and high-density plasmas, and refine accordingly our plasma model.
Firstly, we consider the case of a low density (underdense) plasma, which can be generated via the interaction of a strong optical laser with a thin target. The plasma generation process typically evolves in two steps \cite{FuchsNP2006}: (i) a preplasma is formed by the prepulse of the laser; (ii) this preplasma is subsequently heated by the main laser pulse potentially up to keV electron energies.

We model the plasma following the approach in Refs.~\cite{FuchsNP2006, WuAPJ2017}. With the help of the so-called scaling law, the electron temperature is given as  $T_{\text{e}} \approx 3.6 I_{16} \lambda_{\mu}^2$ keV, where $ I_{16}$ is the laser intensity in units of $10^{16}$ W/cm$^2$ and $\lambda_{\mu}$ the wavelength in microns \cite{Brunel1987, BonnaudLPB1991, GibbonPPCF1996}.
The electron density can be estimated as $n_{\text{e}} = N_{\text{e}}/V_{\text{p}}$ where $N_{\text{e}}$ is the total number of electrons and  the plasma volume is given by $V_{\text{p}} = \pi R_{\text{focal}}^2 d_{\text{p}}$, with  $R_{\text{focal}}$  the focal radius of the laser and the plasma thickness $d_{\text{p}} = c \tau_{\text{pulse}}$ determined by the speed of light $c$ and the laser pulse duration $\tau_{\text{pulse}}$. The electron number can be related to the absorbed laser energy $f E_{\text{pulse}}$ via  $N_{\text{e}} = f E_{\text{pulse}}/T_{\text{e}}$.
Since experimental results in Refs.~\cite{PingPRL2014, PricePRL1995} show that the laser absorption is almost independent of the target material and thickness, we adopt an universal absorption coefficient $f=f(I, \lambda)$ which is a cubic interpolation to theoretical results based on a Vlasov-Fokker-Planck code presented in Ref.~\cite{PingPRL2014}. For the considered intensity range between $10^{15}$ and $2\times 10^{16}$ W/cm$^2$, the absorption fraction $f$ lies between 0.1 and 0.2.

%%%%%%%%%%%%%%%%%%%%%%%%%%%%%%%%%%%
\begin{figure}
\centering
\includegraphics[width=1.0\linewidth]{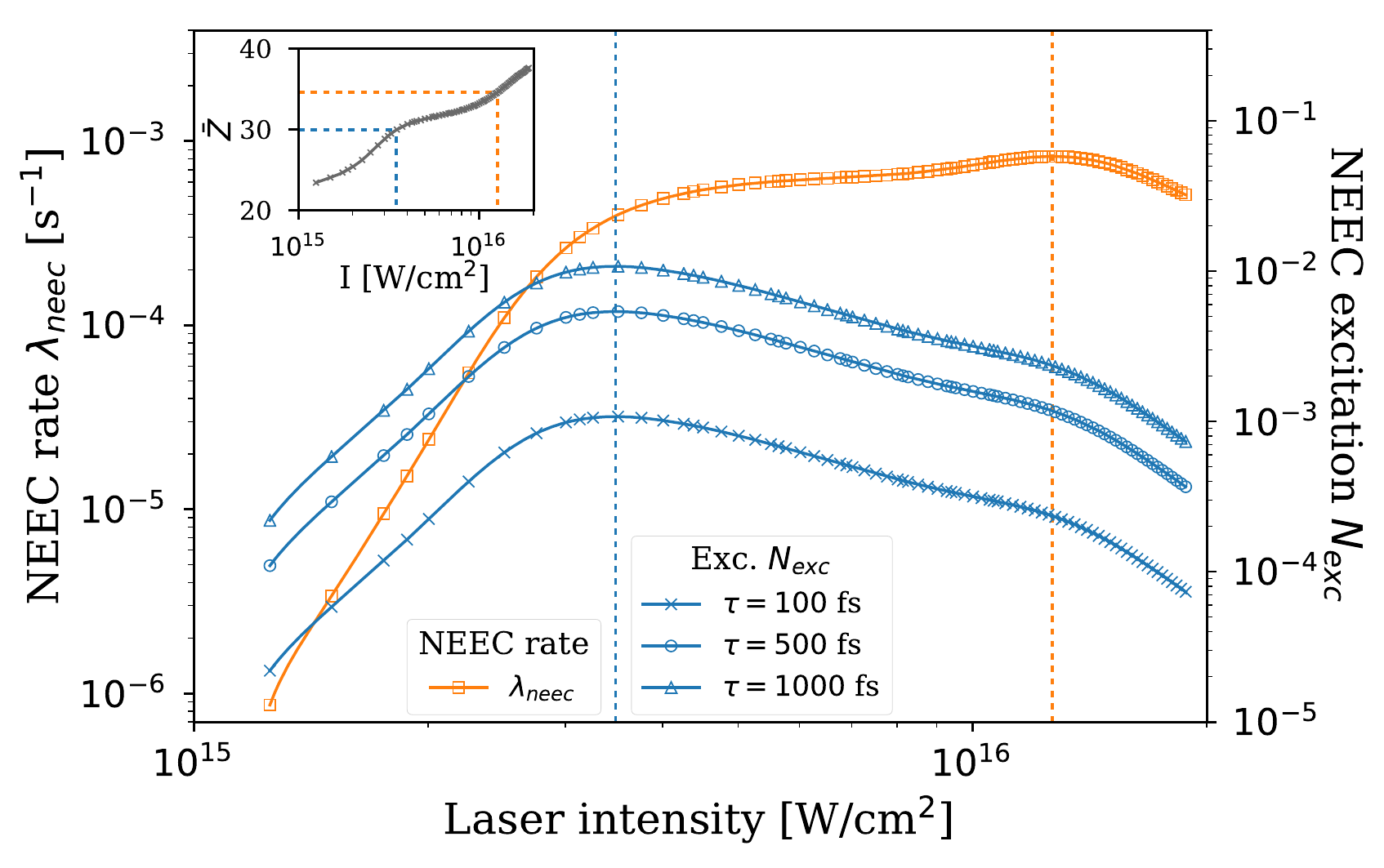}%
\caption{(color online). The NEEC rate $\lambda_{\text{neec}}$ and the total excitation number $N_{\text{exc}}$ per laser pulse  as functions of laser intensity.  The inset shows the average charge state $\bar{Z}$ calculated with the help of FLYCHK. See text for further explanations.
\label{fig:low}}
\end{figure}
%%%%%%%%%%%%%%%%%%%%%%%%%%%%%%%%%%%

For the case of focal radius,  plasma thickness and plasma radius of similar scale, we may again consider the spherical plasma model with the lifetime $\tau_{\text{p}}$.
We use the smallest length scale out of $R_{\text{focal}}$ and $d_{\text{p}}$ to calculate $\tau_{\text{p}}$ for a lower-limit estimate of the NEEC excitation.
Numerical  results for  $\lambda_{\text{neec}}$ and for the total excitation number $N_{\text{exc}}$ per laser pulse  are presented in
Fig.~\ref{fig:low} as a function of the laser intensity. We consider a pulse energy of 100 J, wavelength of 1053 nm, and laser pulse duration values of 100, 500 and 1000 fs, respectively. Also here the optimal laser intensities $I_{\text{opt}}$ at which $\lambda_{\text{neec}}$ and respectively $N_{\text{exc}}$ are maximal do not coincide. For the assumed laser parameters, $\lambda_{\text{neec}}$ is maximized by
$I_{\text{opt}} = 1.3\times 10^{16}$ W/cm$^2$ at a temperature of 5.1 keV and a density of $5.9\times 10^{19}$ cm$^{-3}$. In contrast, the optimal intensity for $N_{\text{exc}}$ per laser pulse is  $3.5\times 10^{15}$ W/cm$^2$  independent of the laser pulse duration in the range of the considered values. The electron temperature and density achieved at this intensity are 1.4 keV and $6.6\times 10^{19}$ cm$^{-3}$, respectively, leading to a charge state distribution with $\bar{Z} \sim 30$
(see inset of Fig.~\ref{fig:low}) where capture channels into the $M$ shell still exceed the $L$-shell contribution.
For $d_{\text{p}}<R_{\text{focal}}$ (the case for the  parameters of Fig. \ref{fig:low}) the plasma lifetime is determined by $d_{\text{p}}$ and in turn by $\tau_{\text{pulse}}$.  The NEEC excitation becomes stronger with increasing laser pulse duration $\tau_{\text{pulse}}$ reaching its maximum at the value where $d_{\text{p}}=R_{\text{focal}}$. For even longer pulse durations we need to use  $R_{\text{focal}}$ in our model to determine the plasma lifetime and this leads to a decrease of $\lambda_{\text{neec}}$.

In Table \ref{table:PWlasers} we evaluate the optimal laser intensity $I_{\text{opt}}$ and  the expected maximal NEEC excitation $N_{\text{exc}}$ for realistic parameters of high-power optical lasers which are  currently available or under construction.
The excitation $N_{\text{exc}}$ per laser pulse is up to six orders of magnitude larger  than the one [$\sim 10^{-6}$, recalculated for the parameters considered here] in the XFEL-generated cold ($T$=350 eV) plasma  \cite{GunstPRL2014, GunstPOP2015}.
The largest value of $1.9$ excitations per pulse should be reached with the PETAL laser which provides both high laser power and long pulse duration.

%%%%%%%%%%%%%%%%%%%%%%%%%%%%%%55

\renewcommand{\arraystretch}{1.25}
\begin{table}
  \centering
  \footnotesize
  \begin{tabular}{lcccc}
  \hline\hline
  & & & & \tabularnewline[-0.4cm]
  & ELI-beamlines & PETAL & LULI & VULCAN \tabularnewline
  & & & & \tabularnewline[-0.4cm] \hline
  & & & & \tabularnewline[-0.4cm]
  $E_{\text{pulse}}$ [J] &1500 & 3500 & 100 & 500  \tabularnewline
  $\tau_{\text{pulse}}$ [fs] &150 & 5000 & 1000 & 500 \tabularnewline
  %$P_{\text{laser}}$ [PW] & 10 & 0.7 & 0.1 & 1 \tabularnewline
  $\lambda$ [nm] & 1053 & 1053 & 1053 & 1053 \tabularnewline
  & & & & \tabularnewline[-0.4cm] \hline
  & & & & \tabularnewline[-0.4cm]
  %$I_{\text{opt}}$ [W/cm$^2$] & $3.5\times10^{15}$ & $3.5\times10^{15}$ & $3.5\times10^{15}$ & $3.5\times10^{15}$ \tabularnewline
  %$n_{\text{e}}$ [cm$^{-3}$] & $6.6\times10^{19}$ & $6.6\times10^{19}$ & $6.6\times10^{19}$ & $6.6\times10^{19}$ \tabularnewline
  %$T_{\text{e}}$ [keV] & 1.4 & 1.4 & 1.4 & 1.4 \tabularnewline
  %$\lambda_{\text{neec}}$ [s$^{-1}$] & $4.0\times10^{-4}$ & $4.0\times10^{-4}$ & $4.0\times10^{-4}$ & $4.0\times10^{-4}$ \tabularnewline
  $N_{\text{exc}}$ & $2.4\times10^{-2}$ & $1.9$ & $1.1\times10^{-2}$ & $2.7\times10^{-2}$ \tabularnewline
  & & & &  \tabularnewline[-0.4cm] \hline\hline
  \end{tabular}
  \caption{Laser parameters and maximal $N_{\text{exc}}$ achieved at the optimal laser intensity $I_{\text{opt}} = 3.5\times 10^{15}$ W/cm$^2$ for  ELI-beamlines L4 \cite{ELI-BL-web}, PETAL \cite{PETAL-web, CasnerHEDP2015}, LULI \cite{LULI-web} and VULCAN \cite{Vulcan-web} lasers. }
  \label{table:PWlasers}
\end{table}
\renewcommand{\arraystretch}{1}

%%%%%%%%%%%%%%%%%%%%%%%%%%%%%%%%

% details on high-density case, PIC

We now turn to the case of high electron densities, which promises the strongest nuclear excitation  according to Fig.~\ref{fig:general}. Experiments and simulations have shown that it is possible to isochorically heat targets at solid-state density to temperatures of a few hundred eV or even a few keV \cite{SaemannPRL1999, AudebertPRL2002, SentokuPOP2007}. Since in this regime the heating of the target is mainly conducted by secondary particles, i.e., hot electrons generated in the laser-target interaction, a  more sophisticated model is necessary  compared to the low-density case.
We have performed a one-dimensional (1D)  particle-in-cell (PIC) simulation of a Nb solid target with 1 $\mu$m thickness and Nb density of $n_{\text{nb}} = 5.5 \times 10^{22}$ cm$^{-3}$ interacting with a high-power laser using the EPOCH code \cite{ArberPPCF2015}. The isomer fraction of $10^{-5}$ is small enough to be neglected here in the determination of the plasma conditions.
The laser has a Gaussian profile in time with peak intensity $I = 10^{18}$ W/cm$^2$,  laser duration  $\tau_{\text{pulse}} = 500$ fs, and laser wavelength $\lambda = 800$ nm, respectively.
A linear preplasma with the thickness of $0.5$ $\mu$m is considered in front of the solid target.  Ionization is not included explicitly in the simulation;  as a representative  order of the electron density, we  fix the charge state to 10.

\begin{figure}
\centering
\includegraphics[width=1.0\linewidth]{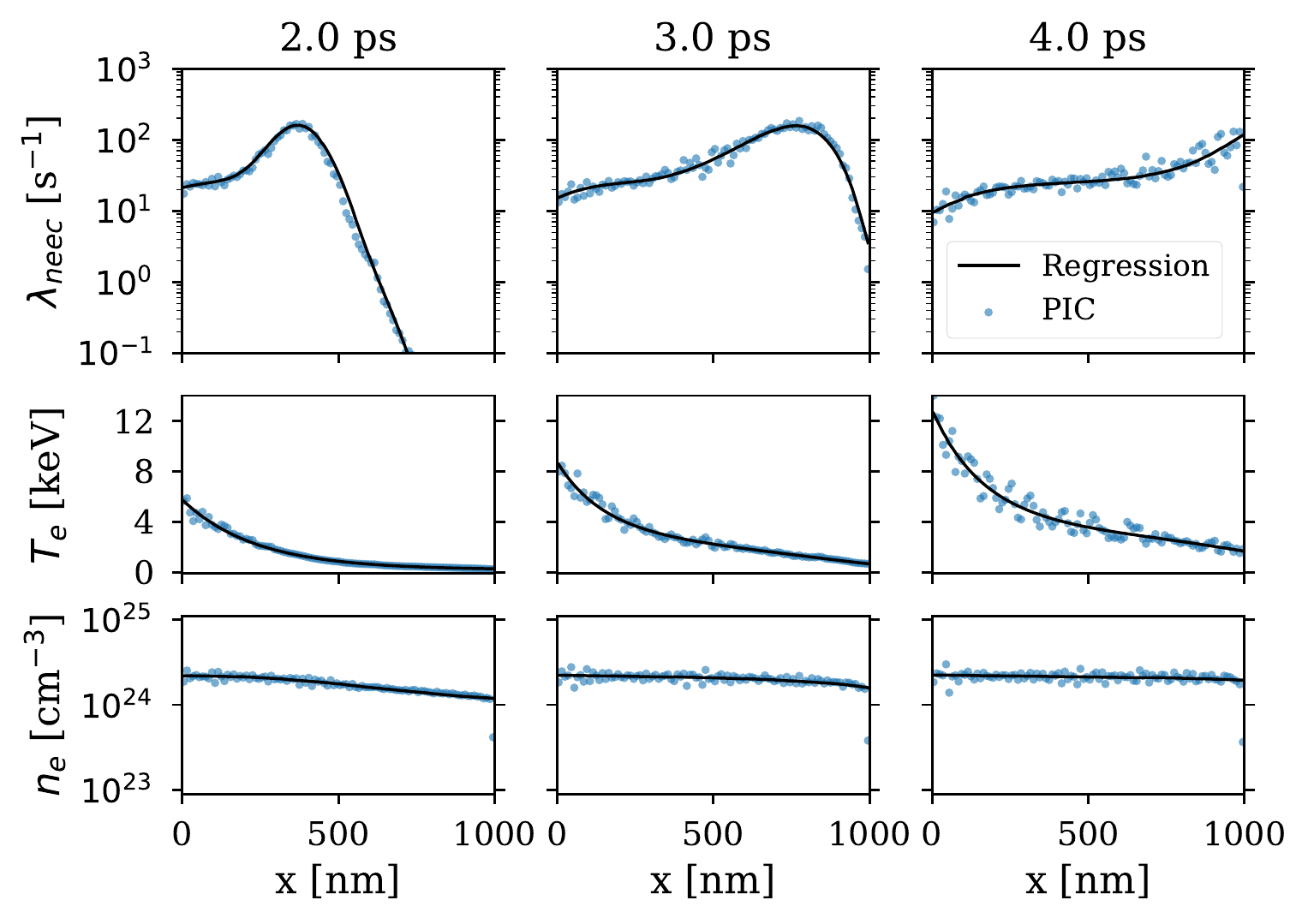}%
\caption{(color online). Electron density, temperature and the NEEC rate based on the PIC simulation as functions of target depth $x$. The laser has peak intensity $I = 10^{18}$ W/cm$^2$ and $\lambda = 800$ nm wavelength. The raw data averaged over 10 nm intervals is presented together with a linear polynomial and a third order exponential fit for $n_{\text{e}}$ and $T_{\text{e}}$, respectively. Regression curves for $\lambda_{\text{neec}}$ calculated with the fitted $n_{\text{e}}$ and $T_{\text{e}}$ functions are shown in the upper graphs.
\label{fig:pic}}
\end{figure}

To include the effect of atomic ionization and recombination events, we averaged the raw data for electron temperature $T_\text{e}$ and ion density $n_{\text{i}}$ from the PIC simulation over 10 nm intervals, and used these values as input for the radiative-collisional model implemented in FLYCHK \cite{FLYCHK2005} to obtain charge state distributions and (corrected) electron densities.
The electron density and temperature values are shown in the lower and middle panels of Fig.~\ref{fig:pic} for the time instants 2, 3 and 4 ps as a function of the target penetration depth $x$.

%%%%%%%%%%%%%%%%%%%%%%%%%5
%%%%%%%%%%%%%%%%%%%%%%%%%%

For  the high-density region, we evaluate the NEEC rate as a function of target depth $x$ and time $t$ by inserting the PIC-simulation results for $T_{\text{e}}$ and the corrected $n_{\text{e}}$ values into Eq.~\eqref{eq:lambda}. The plasma is assumed to be homogeneous only in the plane perpendicular to the $x$ direction over the region of $A_{\text{focal}}$.
We consider a laser pulse energy of 100 J, which leads for the pulse duration and laser intensity adopted in the PIC simulation to a focal spot area of approximatively $2\times 10^{-4}$ cm$^2$.
Results for $\lambda_{\text{neec}}$ are presented in the upper panel of Fig.~\ref{fig:pic}.
The rate is maximized at depths $x$ with optimal plasma conditions for NEEC. The peak propagates through the target and disappears at around 4 ps as  target heating leads afterwards to temperatures exceeding the optimal value. The analysis of the data sampled from 1 to 4 ps in 100-fs steps shows that the integrated NEEC rate reaches its maximum at 3.1~ps and drops roughly to half its value at 4~ps.

Using the regression curves for $\lambda_{\text{neec}}$ calculated with the fitted $n_{\text{e}}$ and $T_{\text{e}}$ functions, we solve Eq.~\eqref{eq:N_exc} in a two-step procedure to obtain $N_{\text{exc}}$.
First, for each time instant $t$ the product of NEEC rate and isomer density is integrated with respect to $x$ over the whole target thickness $d_{\text{t}}$ and multiplied by the focal spot area $A_{\text{focal}}$ to account for the perpendicular directions.
Second, the outcomes of the spatial integration are interpolated as a function of time leading to $N_{\text{exc}}(t)$  which is then inserted into the time integral in Eq.~\eqref{eq:N_exc}. For $t>4$ ps, we extrapolate $N_{\text{exc}}(t)$  assuming an exponential functional behavior initially following the slope at 4 ps. The time integration converges approximatively after 10 ps, leading to an excitation number of $1.8$ isomers per pulse via NEEC which is almost identical with the best value at low densities obtained with the PETAL parameters. With laser repetition rates of few Hz for 100 J pulses,  the threshold of one isomer depletion per second  should be reached providing a detectable signal. The experimental signature of the nuclear excitation would be a  gamma-ray photon of approx.~1~MeV  released in the decay cascade of the triggering level in $^{93}$Mo. An evaluation of the plasma black-body and bremsstrahlung radiation spectra at this photon energy shows that the signal-to-background ratio is very high. Notable here is that in the high-density case a 100 J laser available at many facilities around the world is competitive with a kJ-laser facility.

Tailoring the plasma conditions for NEEC promises a 12 orders of magnitude increase of the $^{93\mathrm{m}}$Mo depletion compared to the direct driving of the nuclear transition with an XFEL laser. 
An experimental proof of this scenario appears to be possible with present high-power optical lasers. The PIC simulation has been carried out in the direction with the smallest length scale of the plasma. Modeling the expansion in the perpendicular direction of the laser incidence, a roughly 10 to 100-times longer plasma lifetime can be expected to boost $N_{\text{exc}}$. A further enhancement can be achieved by employing a combination of optical and x-ray lasers as envisaged for instance at HIBEF \cite{hibef} at the European XFEL \cite{europeanXFEL}.
X-rays-generated  inner shell holes could then provide the optimal capture state independently from the hot plasma conditions.  We note however that further substantial improvements are required for practical energy storage applications. In our calculation, only a $10^{-10}$  fraction of the isomers  in the plasma volume are depleted.  In addition, the total isomer energy stored in the microscopic plasma volume is still far from typical requirements of macroscopic every-day life applications.

%%%%%%%%%%%%%%%%%%%%%%%%%%%%%%%%%%%%%%%%%%%%%%%%%%%%%%%%%

%\bibliographystyle{apsrev}
\bibliographystyle{apsrev-no-url-issn.bst}
\bibliography{NEECLPrefs}{}

\end{document}